\documentclass[fleqn,usenatbib]{mnras}

\usepackage[T1]{fontenc}

\DeclareRobustCommand{\VAN}[3]{#2}
\let\VANthebibliography\thebibliography
\def\thebibliography{\DeclareRobustCommand{\VAN}[3]{##3}\VANthebibliography}

\usepackage{graphicx}	% Including figure files
\usepackage{amsmath}	% Advanced maths commands
\usepackage{amssymb}	% Extra maths symbols

\title[Image restoration and variability in PDS\,70]{Variable structure in the PDS\,70 disc and uncertainties in radio-interferometric image restoration}

\author[S. Casassus \& M. C\'arcamo]{
Simon Casassus,$^{1,2,3}$\thanks{E-mail: simon@das.uchile.cl}
Miguel C\'arcamo,$^{4,5,6}$
\\
$^{1}$ Departamento de Astronom\'{\i}a, Universidad de Chile, Casilla 36-D, Santiago, Chile\\
$^{2}$ Facultad de Ingenier\'ia y Ciencias, Universidad Adolfo Ib\'a\~nez, Av. Diagonal las Torres 2640, Pe\~{n}alol\'{e}n, Santiago, Chile \\
$^{3}$ Data Observatory Foundation, Chile\\
$^{4}$ Jodrell Bank Centre for Astrophysics, Department of Physics and Astronomy, University of Manchester,\\ ~~Alan Turing Building, Oxford Road, Manchester, M13 9PL, UK \\
$^{5}$ University of Santiago of Chile (USACH), Faculty of Engineering, Computer Engineering Department, Chile\\
$^{6}$ Center for Interdisciplinary Research in Astrophysics and Space Exploration (CIRAS), Universidad de Santiago de Chile\\
}

\date{Accepted XXX. Received YYY; in original form ZZZ}

\pubyear{2015}

\begin{document}
\label{firstpage}
\pagerange{\pageref{firstpage}--\pageref{lastpage}}
\maketitle

\begin{abstract}
  
  %Image restoration allows to convey uncertainties in synthesis
  %imaging, as well as faint signal that may have been missed by model
  %images. New image restoration procedures have been proposed in the
  %recent literature, e.g. in the so-called ``JvM correction''.  In
  %this letter we discuss the JvM correction, and propose a strategy
  %for the alignment and joint imaging of multi-epoch visibility
  %data. We show that the JvM correction exaggerates the actual peak
  %signal-to-noise of the data. For an example application we use the
  %available ALMA data in the PDS\,70 disc. We find that the previously
  %reported mm-point source PDS\,70c is indeed detected in the July
  %2019 data, and that it is connected to the cavity wall by a
  %filament, but that it is absent in data from Dec. 2017. PDS\,70c is
  %thus variable by $42\% \pm 13\% $ over a 1.75\,yr time-span. We also
  %pick up fine structure in the inner disc, such that its peak is
  %offset by $\sim 0\farcs04$ from the disc center.  The inner disc is
  %variable too, which we tentatively ascribe to Keplerian rotation as
  %well as intrinsic morphological changes.

  The compact mm-wavelength signal in the central cavity of the
  PDS\,70 disc, revealed by deep ALMA observations, is aligned with
  unresolved H$\alpha$ emission, and is thought to stem from a
  circum-planetary disc (CPD) around PDS\,70c.  We revisit the
  available ALMA data on PDS\,70c with alternative imaging strategies,
  and with special attention to uncertainties and to the impact of the
  so-called ``JvM correction'', which is thought to improve the
  dynamic range of restored images.  We also propose a procedure for
  the alignment and joint imaging of multi-epoch visibility data. We
  find that the JvM correction exaggerates the peak signal-to-noise of
  the data, by up to a factor of 10. In the case of PDS\,70, we
  recover the detection of PDS\,70c from the July 2019 data, but only
  at 8\,$\sigma$.  However, its non-detection in Dec. 2017 suggests
  that PDS\,70c is variable by at least $42\% \pm 13\% $ over a
  1.75\,yr time-span, so similar to models of the H$\alpha$
  variability.  We also pick up fine structure in the inner disc, such
  that its peak is offset by $\sim 0\farcs04$ from the disc
  centre. The inner disc is variable too, which we tentatively ascribe
  to Keplerian rotation as well as intrinsic morphological changes.

\end{abstract}


%%%%%%%%%%%%%%%%%%%%%%%%%%%%%%%%%%%%%%%%%%%%%%%%%%

%%%%%%%%%%%%%%%%% BODY OF PAPER %%%%%%%%%%%%%%%%%%

\section{Introduction}

%explain that, in order to use restored images to

%, it is necessary that both the dirty map of the visibility residuals
%and the smoothed model image bear the same units

%The accretion of giant planets is thought to be regulated by a
%circum-planetary disk (CPD). As summarized in
%Fig.\,\ref{fig:CPDtheory}, this CPD is fed by intricate flows inside
%planetary wakes that cross the annular gap evacuated from the
%circumstellar disc by the
%protoplanet\cite{Szulagyi2020ApJ...902..126S}.  The observation of CPDs would bring crucial information to constrain
%the initial and boundary conditions for giant planet accretion. 

The accretion of giant planets in circum-stellar discs is thought to
be regulated by a circum-planetary disc (CPD), which is fed by
intricate flows inside planetary wakes
\citep[e.g.][]{Szulagyi2020ApJ...902..126S}. However, CPD detections
have been remarkably elusive, except for the compact ro-vibrational CO
signal detected in the central cavity of HD\,100546 by
\citep[][]{Brittain2019ApJ...883...37B}, or the near-IR black-body
component required to account for the spectral energy distribution of
PDS\,70b \citep[][]{Christiaens2019MNRAS.486.5819C}. In general such
CPDs are expected to be faint compact sources, up to one Hill radius
wide and surrounded by complex structures in the parent disc
\citep[e.g.][]{Szulagyi2018MNRAS.473.3573S}. The search for CPDs with
ALMA thus requires bringing the instruments to its limits, and a very
good knowledge of the uncertainties in the resulting images so as to
pick-up genuine signal.

%(~\,au for a 1\,$M_{\rm jup}$ planet orbiting a 1\,$M_\odot$ star at
%100\,au)

Long-baseline ALMA observations of the PDS\,70 disc revealed the
detection of a compact source inside the central
cavity\citep[][]{Benisty2021ApJ...916L...2B}, closely aligned with
unresolved H$\alpha$ signal probably associated to an accreting
protoplanet \citep[i.e. PDS\,70c,][]{Haffert2019NatAs...3..749H}. The
compact mm-wavelength signal may thus stem from a CPD around PDS\,70c
\citep[][]{Benisty2021ApJ...916L...2B}. However, the data analysis
rests on the ``JvM correction'' \citep[][]{JvM1995AJ....110.2037J,
  Czekala2021ApJS..257....2C}, an image restoration procedure that is
thought to improve dynamic range.  Image restoration, applicable to
all synthesis imaging techniques, allows to convey uncertainties, and
to recover faint signal that may have been missed by model images.

  %Image restoration allows to convey uncertainties in synthesis
  %imaging, as well as faint signal that may have been missed by model
  %images. New image restoration procedures have been proposed in the
  %recent literature, e.g. in the so-called ``JvM correction''.  In
  %this letter we discuss the JvM correction, and propose a strategy
  %for the alignment and joint imaging of multi-epoch visibility
  %data. We show that the JvM correction exaggerates the actual peak
  %signal-to-noise of the data. 

The direct detection of a CPD, thanks to deep ALMA observations, is a
momentous achievement in planetary sciences, whose relevance motivates
us to revisit these data with alternative synthesis imaging
tools. Here we perform non-parametric image synthesis using the 
{\sc uvmem} package \citep[][]{Casassus2006,
  Carcamo2018A&C....22...16C} as an alternative to the more commonly
used {\sc clean} algorithm \citep[][]{Hogbom1974A&AS...15..417H,
  RauCornwell2011A&A...532A..71R}.  We give special attention to the
alignment of multi-epoch visibility data, and to the impact of the JvM
correction.

Our approach to the alignment of the multi-epoch visibility data is
described in Sec.\,\ref{sec:alignment}, and is applied to {\sc uvmem}
imaging of PDS\,70 in Sec.\,\ref{sec:PDS70}.  Sec.\,\ref{sec:conc}
summarizes our results. Appendix\,\ref{sec:IS} describes the standard
techniques of image restoration and visibility gridding, before
showing that the JvM correction exaggerates the sensitivity of
restored images.

\section{Alignment of visibility data} \label{sec:alignment}

% or simply consecutive observations with
% different correlator settings,
 
Datasets obtained with different source acquisitions, i.e.
multi-epoch observations, may have different absolute flux scales or
be affected by different pointing errors. For instance, the flux
calibration accuracy for ALMA is thought to be around $\sim$10\%, and
the phase center accuracy (absolute positional accuracy) is, very
approximately, about 1/10 the size of the clean beam major axis
\citep[see for
instance][]{ALMA_technical_handbook2019athb.rept.....R}. In order to
restore images using concatenated datasets, it is necessary to align
them. It is also necessary to scale them in flux, as otherwise
different flux calibrations will lead to stronger residuals.

%10.5.2 of ALMA_technical_handbook2019athb.rept.....R

Here we propose to align such datasets in the $uv-$plane, assuming
that the source is not variable in time. The result of our procedure
is similar to that of joint self-calibration (described in
Sec.\,\ref{sec:PDS70imaging}), but it restricts the number of unknowns
to only a single position shift and a flux scale factor, while
self-calibration adjusts time-series for the antenna gains. Another
difference is that  self-calibration requires a bright source to define an adequate
self-calibration model. We describe our procedure for two datasets,
labeled $S$ and $L$ anticipating the application to PDS\,70, where the
datasets correspond to different array configurations, with shorter
and longer baselines\footnote{this alignment procedure can of course
  be applied to datasets acquired with the same configuration}. We
extract subsets of each dataset with exactly the same spectral domains
in LSRK. We then gridd each dataset to a common $uv-$grid, in natural
weights, so according to Eqs.\,\ref{eq:gridV} and \ref{eq:wadjust_nat}
(which are implemented in the {\sc pyralysis}\footnote{see Data Availability} software package). This
results in visibilities $\left\{\tilde{V}^S_k\right\}_{k=1}^N$ and
$\left\{\tilde{V}^L_k \right\}_{k=1}^N $, for the short and long
baseline datasets, respectively. The common $uv$-grid is set by the
shortest and longest baselines. We then aligned each pair by shifting
$\tilde{V}^S$, i.e. by minimising
\begin{equation}
  \chi^2_{\rm align}(\alpha_R, \delta{\vec{x}}) = \sum_{k=1}^N W^{\rm align}_k \| \tilde{V}_k^L - \tilde{V}_k^{Lm}   \|^2,
\end{equation}
where
\begin{equation}
  \tilde{V}_k^{Lm} = \alpha_R \, e^{i 2\pi \,\delta\vec{x}\cdot\vec{u}_k} \, \tilde{V}_k^S,
\end{equation}
corresponds to a spatial shift in origin by $\delta \vec{x}$ and to
a scaling of the absolute fluxes for $\tilde{V}^S$ by  $\alpha_R \in \Re$. The weights are
\begin{equation}
  W_k^{\rm align} = \frac{W_k^S \, W_k^L}{W_k^S+  W_k^L}, \label{eq:commonW}
\end{equation}
where $W^S_k$ and $W^L_k$ are each given by Eq.\,\ref{eq:Wk}, for both
datasets.  The overlap in $uv$-radii, over which the optimisation was
carried out, was determined by taking azimuthal averages in the
gridded weights $W^S$ and $W^L$, and selecting the range where the
radial profiles are both above 1/10 their peak.  See
Sec.\,\ref{sec:PDS70imaging} below for an example application and for
independent checks on the alignment procedure. This visibility
alignment procedure is implemented in the package {\sc VisAlign}\footnote{see Data Availability}.

\section{Application to PDS\,70} \label{sec:PDS70}

\subsection{Imaging} \label{sec:PDS70imaging}

We refer to \citet[][]{Long2018ApJ...858..112L},
\citet[][]{Keppler2019A&A...625A.118K},
\citet[][]{Isella2019ApJ...879L..25I} and
\citet[][]{Benisty2021ApJ...916L...2B} for a description of the
observations. We followed the same procedures for the extraction of
the continuum visibility data, and adopt the same nomenclature as in
\citet[][]{Benisty2021ApJ...916L...2B} to refer to the various
datasets. In summary, we used the datasets corresponding to the
following ALMA programmes: {\tt 2015.1.00888.S}, from August 2016,
with short baselines (15\,m to 1.5\,km) and referred to as SB16; {\tt
  2017.A.00006.S}, from Dec. 2017, with intermediate baselines (15\,m
and 6.9\,km) and referred to as IB17; and {\tt 2018.A.00030.S}, from
July 2019, with long baselines (92\,m to 8.5\,km) and referred to as
LB19.

%The available ALMA data-set on PDS\,70,
%from \cite{Long2018ApJ...858..112L, Keppler2019A&A...625A.118K,
%  Isella2019ApJ...879L..25I, Benisty2021ApJ...916L...2B}, is
%(following the nomenclature of \cite{Benisty2021ApJ...916L...2B}):
%{\tt 2015.1.00888.S}, from August 2016, with short baselines (15\,m to
%1.5\,km) and referred to as SB16; {\tt 2017.A.00006.S}, from
%Dec. 2017, with intermediate baselines (15\,m and 6.9\,km) and
%referred to as IB17; and {\tt 2018.A.00030.S}, from July 2019, with
%long baselines (92\,m to 8.5\,km) and referred to as LB19.
%The detection of signal coincident with PDS\,70c is confirmed
%(Fig.\,\ref{fig:PDS70_concats_all}a), with point-like signal reaching
%126\,$\mu$Jy\,beam$^{-1}$, or 8\,$\sigma$. This point source is
%surrounded by extended emission, in the form of a filament or spur,
%that connects with the cavity wall.  The color map in
%Fig.\,\ref{fig:PDS70_concats_all} is chosen to higlight the faint
%features and the noise, while also tracing the brighter features. The
%noise level in Fig.\,\ref{fig:PDS70_concats_all}c being
%31.9\,$\mu$Jy\,beam$^{-1}$, a 126\,$\mu$Jy\,beam$^{-1}$ point source
%should have been picked up at 4$\sigma$.

Provided with the continuum visibility data, we proceeded to
self-calibrate \citep[e.g.][]{1999ASPC..180..187C} each dataset
independently. For bright sources this procedure  improves
calibration and may allow reaching thermal noise in the restored image.
Given that \citet[][]{Benisty2021ApJ...916L...2B} used the JvM
correction, in order to set a comparison point before investigating
alternative imaging strategies, we first applied the standard
self-calibration procedure, as implemented in {\sc casa}, and composed
of successive iterations of the tasks {\tt tclean}, {\tt gaincal} and {\tt
  applycal}. We combined all spectral windows (spws), 
proceeding as in \citet[][]{Benisty2021ApJ...916L...2B}, and reached a 
peak signal to noise ratio (PSNR) in the final restored images of
36.3 using Briggs weighting with $r=2$.

%This version includes a {\tt clean} surrounding PDS\,70c, as this
%promotes signal in the model image at this location (the {\sc clean}
%algorithm is known to mimic a point source even in the absence of
%signal, if a box is placed on pure noise).  ***MIGUEL WE NEED TO SHOW
%THE VERSION WITHOUT THIS CLEAN BOX ON PDS70C. IT IS NOT INCLUDED IN
%THE TAR BALL THAT YOU SHARED***

As an alternative imaging strategy, we used the {\sc uvmem} package to
replace the {\tt tclean} model in the {\sc casa} selfcal loop
\citep[see][for more details on {\sc uvmem} and recent applications to
ALMA data of protoplanetary discs]{Casassus2018MNRAS.477.5104C,
  Casassus2019MNRAS.483.3278C, Casassus2021MNRAS.507.3789C}. We used
an elliptical mask enclosing the source that restricts the number of
free parameters \citep[as in][who used Keplerian masks for line
data]{Casassus2021MNRAS.507.3789C}. For very bright and compact
sources, as in SB16 or IB17, this mask is redundant in terms of the
final PSNR, but it helps in case of lower dynamic range, as for
LB19. For all datasets, we used a pure least-squared ($\chi^2$)
optimization, regularized only by the requirement of image positivity
(i.e. no entropy term was applied). We then proceeded to
self-calibrate the data using a special purpose programming framework,
{\sc OOselfcal}\footnote{see Data Availability}, which is essentially a wrapper for {\sc casa} tasks
{\tt gaincal} and {\tt applycal} and any choice of imager (in this
case {\sc uvmem}). {\sc OOselfcal} allows execution with as little user
interaction as possible.  This framework starts with phase-calibration
only, in selfcal iterations that progressively lower the solution
interval by half, starting from the scan length (about 50s). These
iterations are interrupted when the PSNR\footnote{In our framework
  this PSNR is calculated as the ratio of peak signal in the restored
  images to the noise in the whole residual image, with a field of
  view of 4\,arcsec in the case of PDS\,70} decreases, and the
self-calibrated visibility dataset is reverted to the calibration
tables that yielded the highest PSNR. The phase-calibration stage is
then followed by a single loop of amplitude and phase
self-calibration, also tested for improvement using the PSNR. For LB19 the
phase-calibration progressed until the first iteration, which was
enough to reach a PSNR of 56 with $r=2$. For SB16, a PSNR of 266
($r=2$)was reached after two phase calibration loops. For IB17, a
single iteration in phase  yielded a PSNR of 75 ($r=2$).

\begin{figure}
  \centering
  \includegraphics[width=\columnwidth,height=!]{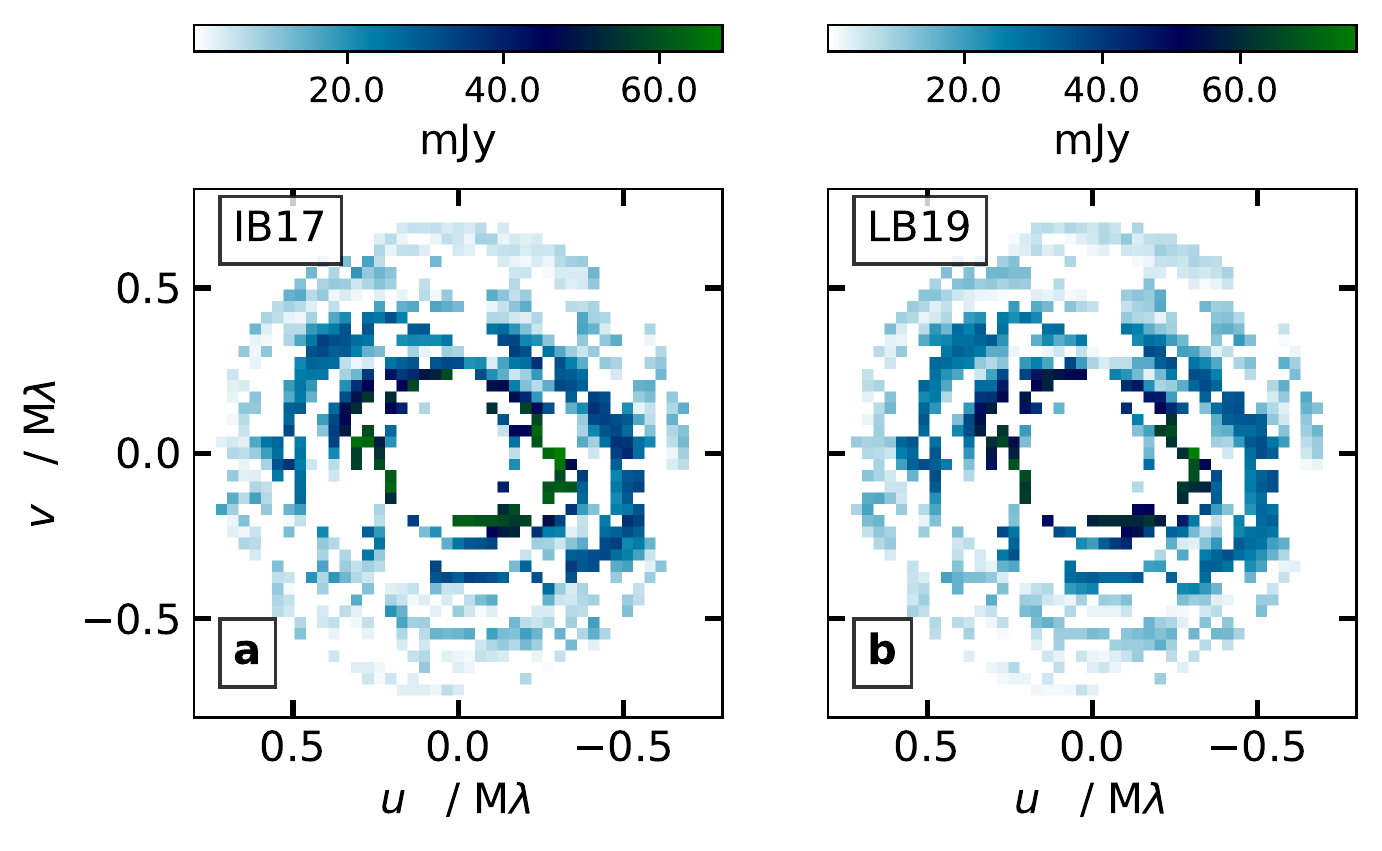}
  \caption{Gridded visibility amplitudes for datasets IB17 (a) and LB19 (b). The data have been filtered for the common $uv-$domain, according to Eq.\,\ref{eq:commonW}.  } \label{fig:powerimages}
\end{figure}

Having self-calibrated each dataset independently, they were
concatenated for joint imaging. This required alignment to a common
reference dataset, for which we chose LB19, both for position and
absolute flux calibration. The alignment was achieved in the
$uv$-plane, using {\sc VisAlign} (as described in
Sec.\,\ref{sec:alignment}). We therefore assume that the bulk of
PDS\,70 is not variable between the two epochs\footnote{Since the bulk
  of the flux density, by far, is dominated by the outer ring,
  Keplerian rotation could trace an arc of $\sim$10\,mas in length,
  subtending an angle of $\sim$0.7\,deg in azimuth relative to the
  star. This is a small angular rotation, that probably does not bias
  the alignment procedure, which fits for a scaling factor and a
  positional translation}. We extracted a frequency domain common to
all 3 datasets, in LSRK, and then aligned each pair SB16/LB19 and
IB17/LB19. The common $uv$-grids correspond to 2048$\times$2048 images
with pixel sizes of  3\,mas and 4\,mas,
respectively. Fig.\,\ref{fig:powerimages} shows the visibility
amplitudes for the grid corresponding to the IB17/LB19 alignment. For
the SB16 and LB19 alignment, the common range in $uv$-radii extended
from $0.17\,M\lambda$ to $0.47\,M\lambda$. The fit yielded
$\delta\vec{x} = (-0\farcs012 \pm 0\farcs001, -0\farcs019\pm0\farcs
001)$. The flux scale factor was $\alpha_R = 0.788\pm0.004$, which is
somewhat smaller than the ratio of flux densities extracted within the
mask, of $0.861 \pm 0.004$. The LB19 flux density is biased upwards in
our pure $\chi^2$ reconstructions, because of the positive-definite
noise in the model images. Using entropy regularization resulted in a
lower flux density ratio of $0.77\pm 0.004$ (but with stronger
synthesis imaging artifacts, which is why we use the pure $\chi^2$
versions). For the IB17 and LB19 alignment,  $uv$-range extended
from $0.17\,M\lambda$ to $0.71\,M\lambda$, $\alpha_R = 1.001\pm 0.002$
and
$\delta\vec{x} = (-0\farcs01401 \pm 0\farcs0003,
-0\farcs0203\pm0\farcs 0004)$. An absolute flux scale difference of
$\sim$20\% between SB16 and IB17 or LB19 is not surprising for ALMA
Band\,7 \citep[][]{Francis2020AJ....160..270F}. The alternative, that
the flux of the disc itself is time-variable, is unlikely since the
flux scales between IB17 and LB19 are consistent, while the time
separations between IB17 and SB16, and between LB19 and IB17, are
similar.

%, and may be due to
%different weather conditions between the acquisitions of the source
%and of the flux calibrator.

%As an example
%application of our alignment procedure we report the comparison
%between SB16 and LB19.

\begin{figure}
  \centering
  \includegraphics[width=\columnwidth,height=!]{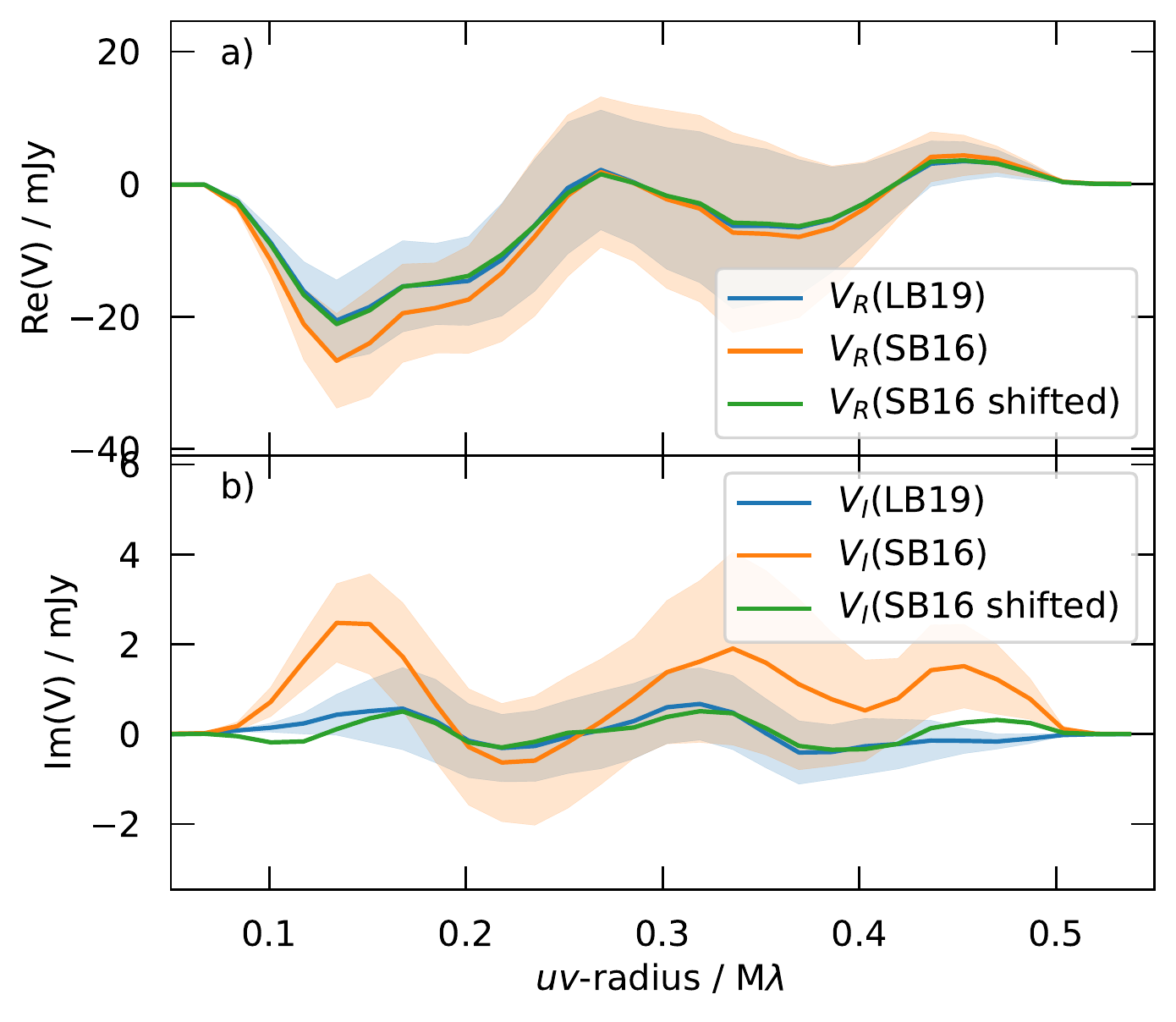}
  \caption{Visibility $uv$-spectra for SB16 and LB19, and $uv$-plane alignment. We plot the deprojected and azimuthally-averaged visibilities of PDS\,70, over a common range in $uv-$radii (see text), with real parts in {\bf a)} and imaginary parts in  {\bf b)}. The total heights of the shaded regions correspond to the standard deviation of azimuthal scatter. We do not plot this scatter for the shifted visibilities to avoid crowding.} \label{fig:powerspectra}
\end{figure}

\begin{figure}
  \centering
  \includegraphics[width=\columnwidth,height=!]{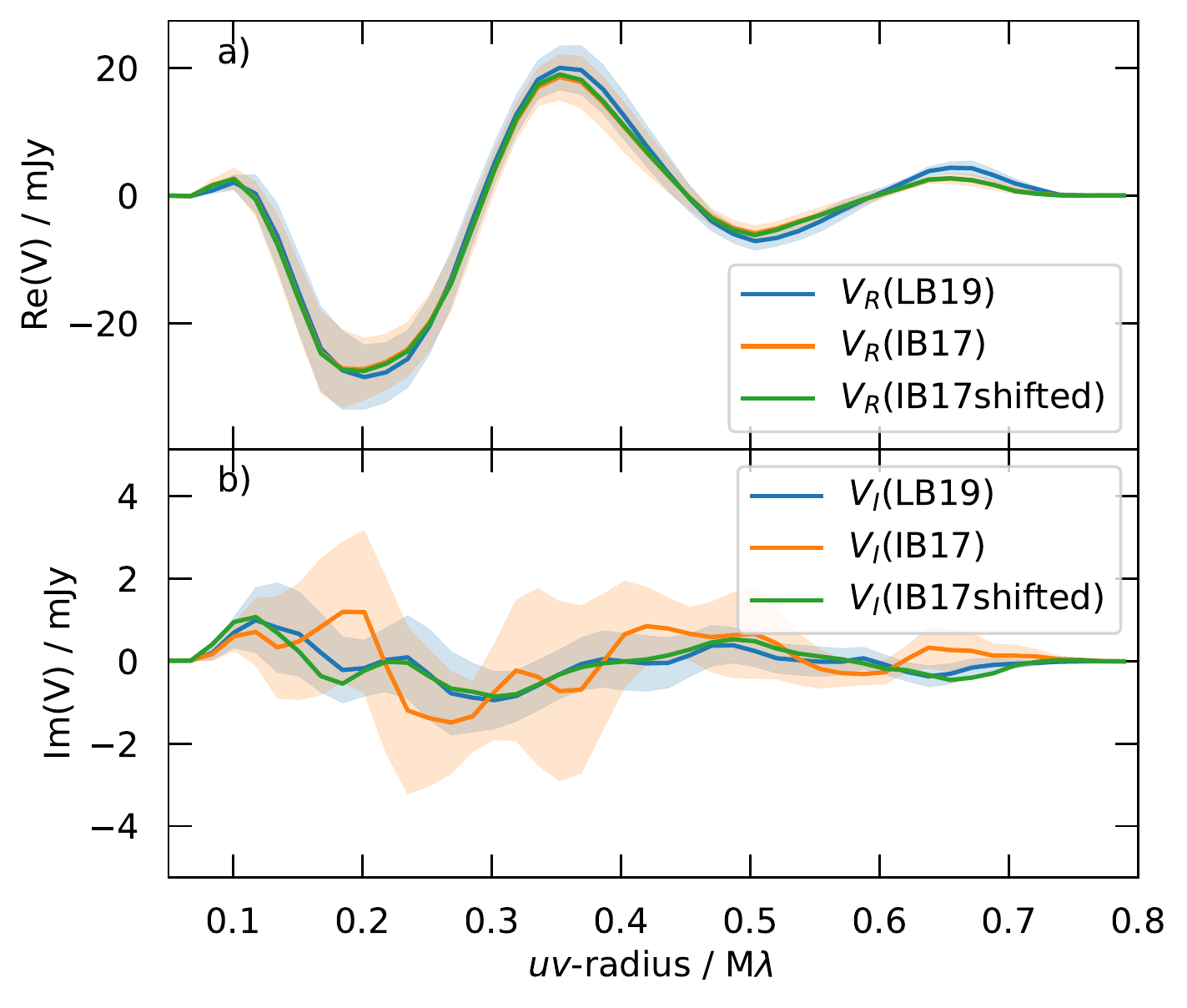}
  \caption{Visibility $uv$-spectra for IB17 and LB19, and $uv$-plane alignment. Annotations follow from Fig.\,\ref{fig:powerspectra}.} \label{fig:powerspectraIB17}
\end{figure}

The impact of the alignment is summarised in
Figs.\,\ref{fig:powerspectra} and \ref{fig:powerspectraIB17}. Since
the source is a fairly axially symmetric disc, albeit projected on the
sky, we rotated and compressed  the $uv$-plane to compensate for the
disc orientation, with an inclination $i=130\,$deg and
$\text{PA}=160.4$\,deg (see below).

%The ranges in abscissa in
%Figs.\,\ref{fig:powerspectra} and \ref{fig:powerspectraIB17} are
%larger compared to the $uv$-radial domain of the {\sc VisAlign}
%because of the deprojection (i.e. the projection of  baselines onto
%the sky plane leads to shorter norms in the $uv$-plane).
%

As the disc of PDS\,70 is essentially a 
ring, we can use the center of the ring to test the accuracy of our
alignment procedure. The {\sc MPolarMaps} package
\citep[][]{Casassus2021MNRAS.507.3789C} estimates the disc
orientation, under the assumption of axial symmetry, by minimizing the
azimuthal dispersion of radial intensity profiles. We compare the LB19
image, in its restored version (see below), with the IB17 model image
(see below). For LB19, the best fit orientation corresponds to
PA$=160.4\pm0.08$\,deg, $i=130.0\pm0.06$\,deg, and a ring center
offset by
$\vec{\Delta}_{\rm LB19} = (-0\farcs0018 \pm
0\farcs0003,-0\farcs0036\pm0\farcs0003)$ relative to the phase
center. For IB17, we obtained PA$=160.5\pm0.18$\,deg,
$i=129.9_{-0.14}^{+0.16}$\,deg, and a ring center at
$\vec{\Delta}_{\rm IB17} =
(-0\farcs0032^{+0\farcs0006}_{-0\farcs0009},-0\farcs0025\pm0\farcs0008)$
relative to the phase center. The ring centers
$\vec{\Delta}_{\rm LB19}$ and $\vec{\Delta}_{\rm IB17}$ coincide
within $\sim$1\,mas, or $1.5\,\sigma$, in support of the accuracy of
the $uv$-plane alignment with {\sc VisAlign}.

%#LB19
%filename_source="/home/simon/PDS70/pyra/output_OOselfcal_usermask_chi2_casarestore_2inf/_ph0.restored.fits"
%(base) belka14:31:54~/PDS70/polarmaps$python Mextract_profile.py
% PA (160.43462989005818, 0.08052937023182949, 0.08133591186665967) 160.43464768398135
%inc (130.00771070761903, 0.06167397780936312, 0.05594255831630335) 130.0139017966674
%dra_off (-0.0018469868450791999, 0.00028235738769911197, 0.00027775892453939287) -0.0018763881615685017
%ddec_off (-0.0036938855407735293, 0.0003426333536842006, 0.00030544576715543174) -0.003675204338760792

%#IB17
%filename_source='/home/simon/common/ppdisks/PDS70/data/output_OOselfcal_concat_SB16_IB17_apcal/_ph0.fits'
%param     distrib     max 
%PA (160.52622482642235, 0.1872687134650448, 0.18129192504466118) 160.52538251694946
%inc (129.91203675749836, 0.14070851935616702, 0.1577402059882047) 129.89462781500902
%dra_off (-0.0032280241650962465, 0.0006029439975628052, 0.000993180922320896) -0.0032585930594504005
%ddec_off (-0.00254975924997542, 0.0007848344804888823, 0.0008315529559772801) -0.00247754454316747

Having aligned all 3 datasets, we proceeded with joint imaging. We
concatenated the visibility data directly, i.e. with the application
of the {\tt statwt} task in CASA to replace the visibility weights by
those corresponding to the dispersion of each datum, but without
additional modification.  We applied the {\sc OOselfcal} framework,
which yielded improvements for SB16+IB17+LB19 (PSNR=95) and also for
SB16+IB17 (PSNR=94), both  after a single round of phase calibration. The
amplitude and phase self-calibration resulted in an improvement only
for SB16+LB19 (PSNR=84).  A selection of images resulting from the
present imaging procedure is given in
Fig.\,\ref{fig:PDS70_concats_all}.

%alpha_R 0.7888173108036363
%delta_x -0.012974736573359062
%delta_y -0.01890070895096003
%best fit  [0.7888173108036363, -0.012974736573359062, -0.01890070895096003]
%errors   [0.0026870204738300707, 0.0005338399413724274, 0.0006503153662926284]
%bestchi2  3213.113569688698
%red bestchi2  1.9628060902191191
%Hessian errors scaled for red chi2 = 1
%errors   [0.00376452 0.00074791 0.00091109]

%(base) belka17:30:02~/PDS70/pyra$python plot_powerspectra.py
%(base) belka17:29:25~/PDS70/pyra$rsync -va fig_powersprectra_SB16_defaultuvrange_4report.pdf ~/common/ppdisks/PDS70/rep_restor/figs/

%(base) belka17:30:02~/PDS70/pyra$python plot_powerspectra.py
%(base) belka17:29:25~/PDS70/pyra$rsync -va fig_powersprectra_SB16_defaultuvrange_4report.pdf ~/common/ppdisks/PDS70/rep_restor/figs/

% python plot_powerspectra_IB17.py

\subsection{Results and discussion}

\subsubsection{Noise level in the  LB19 image}

The noise level reported in Fig.\,\ref{fig:PDS70_concats_all}a, of 16.4\,$\mu$Jy\,beam$^{-1}$,
appears to match the expected instrumental sensitivity. The ALMA
Observing Tool (OT, provided by the Joint ALMA Observatory to prepare
ALMA proposals) predicts about 6\,h total execution time to reach a
noise level of 17\,$\mu$Jy\,beam$^{-1}$, under default weather
conditions. This seems fairly consistent with the actual total
execution time for the LB19 dataset, of 3.7h, considering that is was
carried out under very good weather conditions (with precipitable water
vapour columns ranging from 0.6 to 0.7\,mm). However, the noise level
quoted by \citet{Benisty2021ApJ...916L...2B} for the LB19 dataset
using Briggs robustness parameter of 2.0 is only
6.2\,$\mu$Jy\,beam$^{-1}$, which according to the OT would require
2\,d. We suggest that this very low noise level is due to the
application of the JvM correction (see Sec.\,\ref{sec:JvM}).

\subsubsection{PDS\,70c}

\begin{figure*}
  \centering
  \includegraphics[width=\textwidth,height=!]{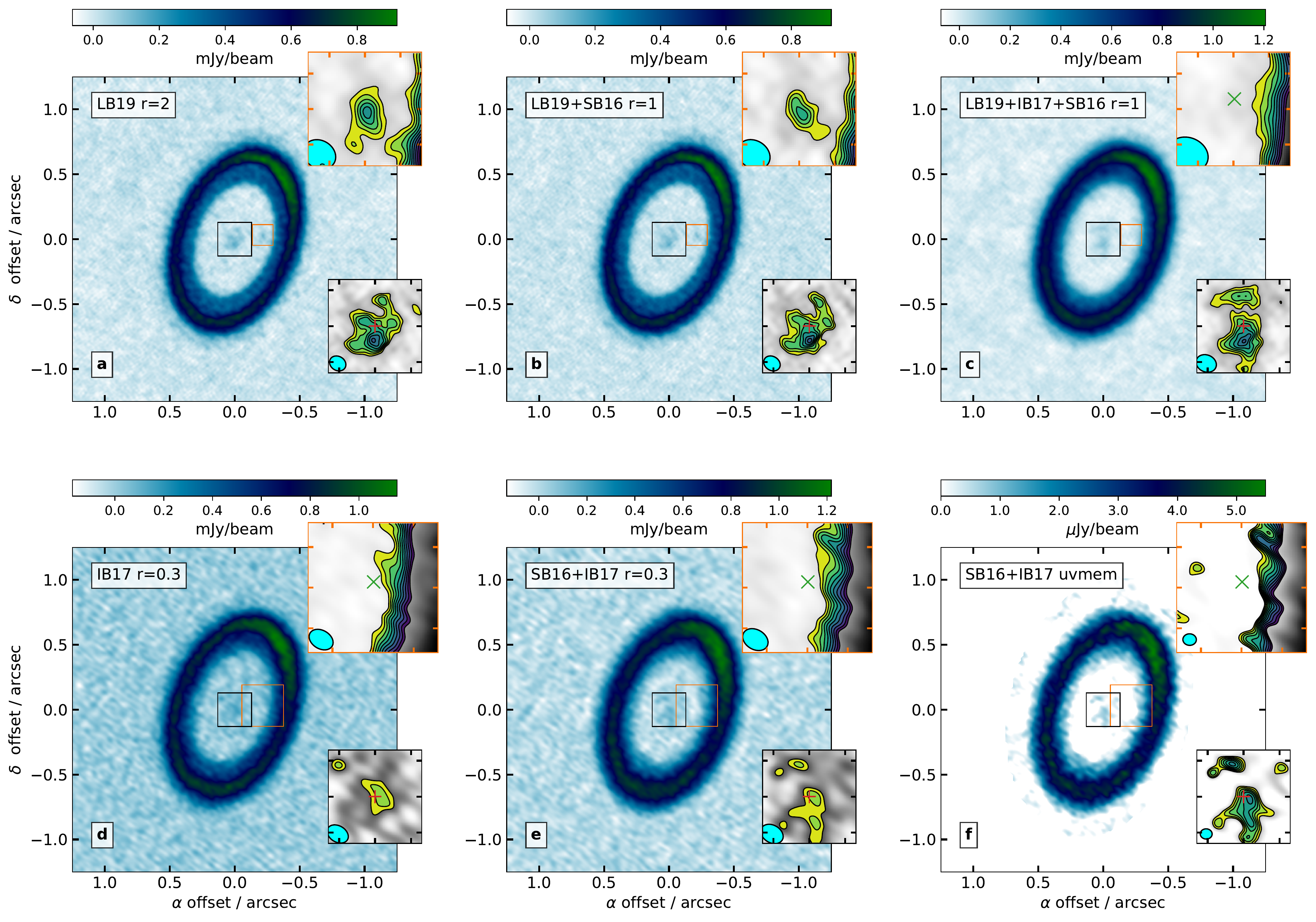}
  \caption{Multi-configuration imaging at 351\,GHz in PDS\,70.  {\bf
      a:} Restored {\sc uvmem} image for LB19, with Briggs robustness
    parameter $r= 2.0$, resulting in a clean beam of
    $\Omega_b = 0\farcs048\times 0\farcs039 \,/\,59$\,deg, where we
    give the beam major axis (bmaj), minor axis (bmin) and direction
    (bpa) in the format bmaj$\times$bmin/bpa. The noise in the
    residual image is $\sigma = 16.4\,\mu$Jy\,beam$^{-1}$. The two
    insets zoom on the central region, with tick-marks separated by
    $0\farcs1$, and on PDS\,70c, with tick-marks at $0\farcs05$. In
    both insets the linear grey scale stretches over the range of
    intensities in each region, and the contour levels start at
    $3\sigma$ and are incremented in units of $\sigma$. Intensities
    within each contour level are color-coded differently.  {\bf b:}
    same as a) but for the concatenation LB19+SB16, with
    $\Omega_b = 0\farcs048\times0\farcs039\,/\,58$deg, and
    $\sigma = 15.7\,\mu$Jy\,beam$^{-1}$ {\bf c:} same as a) but for
    the concatenation LB19+SB16+IB17, with
    $\Omega_b = 0\farcs058\times0\farcs047\,/\,58$deg, and
    $\sigma = 15.8\,\mu$Jy\,beam$^{-1}$.  {\bf d:}
    Same as a) but for IB17, restored with $r=1$, with
    $\Omega_b = 0\farcs078\times 0\farcs064 \,/\,59$\,deg, and
    $\sigma = 29.6\,\mu$Jy\,beam$^{-1}$.  In the inset centered on
    PDS\,70c, the tick marks are separated by $0\farcs1$, and the
    green cross marks the position of PDS\,70 at the epoch of IB17.
    {\bf e:} Same as c) but for SB16+IB17, restored with $r=0.3$, with
    $\Omega_b = 0\farcs067\times 0\farcs049 \,/\,60$\,deg, and
    $\sigma = 31.9\,\mu$Jy\,beam$^{-1}$. {\bf f:} {\sc uvmem} model
    image for SB16+IB17, with an approximate resolution of 1/3 the
    natural weight beam \citep[][]{Carcamo2018A&C....22...16C}, or
    $\Omega_b \approx 0\farcs033\times 0\farcs029 \,/\,88$\,deg. The
    contours start at 3\,$\sigma$, where
    $\sigma = 0.1\,\mu$Jy\,pix$^{-1}$ is a representative noise level.
  } \label{fig:PDS70_concats_all}
\end{figure*}

\begin{figure}
  \centering
  \includegraphics[width=\columnwidth,height=!]{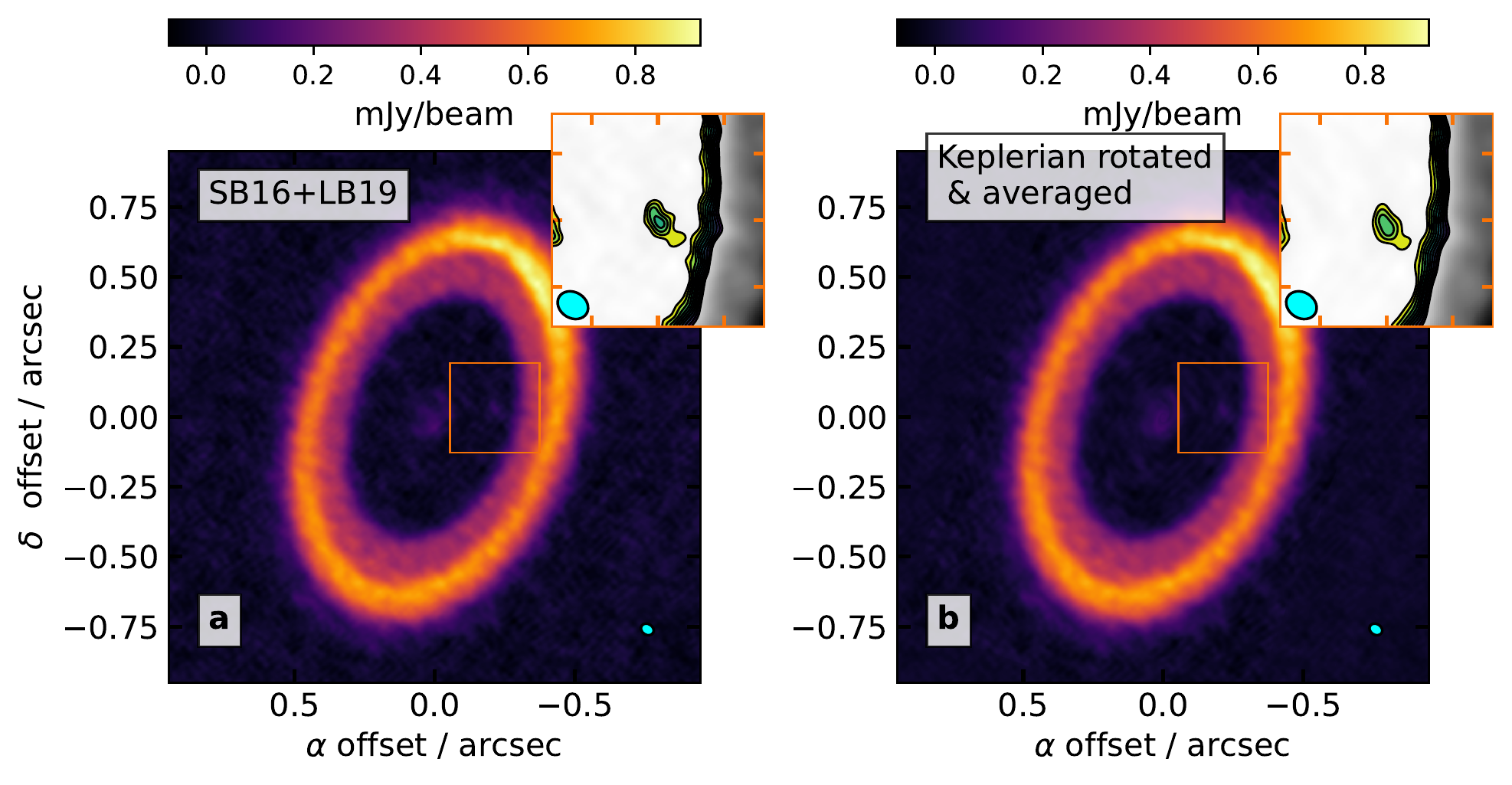}
  \caption{The impact of Keplerian rotation over a time-span of
    1.75\,yr.  {\bf a)} Restored image for LB19+SB16, same as
    Fig.\,\ref{fig:PDS70_concats_all}b but for a different color
    scheme. {\bf b)} Average of the image in a), and its
    transformation to a Keplerian phase rotated by
    1.75\,yr.} \label{fig:keprot}
\end{figure}

As summarised in Fig.\,\ref{fig:PDS70_concats_all}, we confirm the
detection of signal coincident with PDS\,70c. This is best seen in
Fig.\,\ref{fig:PDS70_concats_all}a, with point-like signal reaching
126\,$\mu$Jy\,beam$^{-1}$, or 8\,$\sigma$. We note that this point
source is surrounded by extended emission, in the form of a filament
or spur, that connects with the cavity wall. This spur is best seen in
Fig.\,\ref{fig:PDS70_concats_all}b, where it is detected at
3$\sigma$. The color map in Fig.\,\ref{fig:PDS70_concats_all} is
chosen to higlight the faint features and the noise, while also
tracing the brighter features. A more standard color map is used in
Fig.\,\ref{fig:keprot} \citep[the latter version is easier to compare
with the color coding used by ][and also with ESO press release
eso2111b]{Benisty2021ApJ...916L...2B}.

%genfigs/summary_concats.py

    %Same as a) but for IB17, restored with $r=0.3$, with
    %$\Omega_b = 0\farcs063\times 0\farcs045 \,/\,59$\,deg, and
    %$\sigma = 33.54\,\mu$Jy\,beam$^{-1}$.

%This filament is not entirely detected in the model image
%(Figs.\,\ref{fig:PDS70_concats_LB19}d and \ref{fig:PDS70_concats_LB19}e),
%which shows that it is a faint structure, but nonetheless in the
%residuals. 

%. It is barely picked up in the associated model image in
%Fig.\,\ref{fig:PDS70_concats_all}f

Intriguingly, the point source PDS\,70c is missing in
LB19+IB17+SB16. In order to test whether this is an effect of 
averaging different Keplerian rotation phases, we have de-projected
the LB19+SB16 data, transformed the map by Keplerian rotation over a
period of 1.75\,yr around a 0.9\,$M_\odot$ star, re-projected onto the
sky plane and averaged the result with the initial
image. Fig.\,\ref{fig:keprot} shows that, while PDS\,70c is somewhat
fainter in the average than in the initial image, it is still quite
conspicuous.

%python summary_kepav.py

\begin{figure}
  \centering
  \includegraphics[width=\columnwidth,height=!]{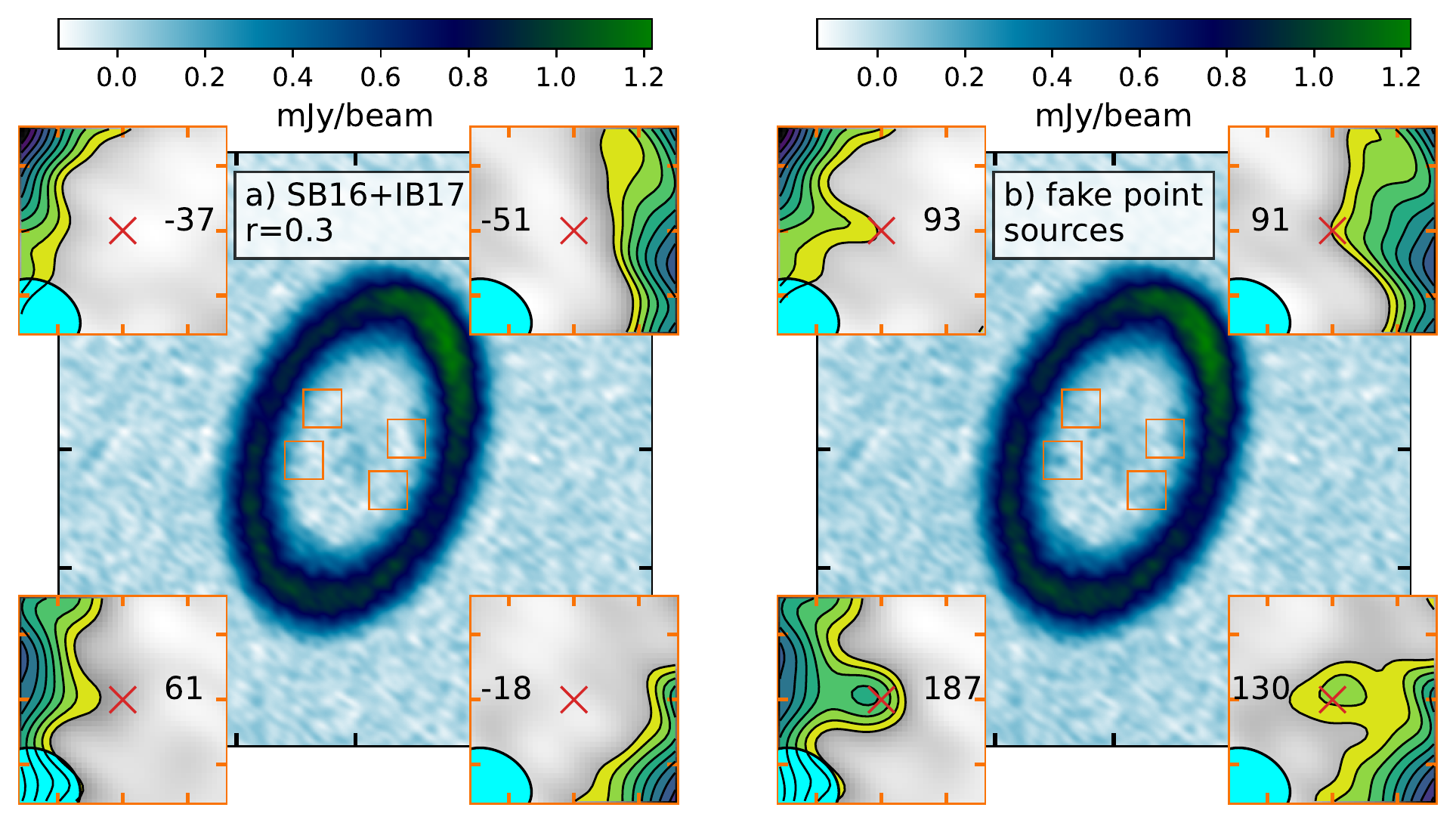}
  \caption{Tests of point source sensitivity through the injection of
    point sources in IB\,17. a): same as
    Fig.\,\ref{fig:PDS70_concats_all}e, but with 4 insets that zoom in
    on the expect position of PDS\,70c, and 3 other locations, at
    symmetrical positions relative to the disc axes. The red crosses
    mark the centre of each inset, and the numbers close to the red
    crosses correspond to the specific intensity value at the
    cross. b): same as a), but with the addition of fake 126\,$\mu$Jy
    point sources at all 4 locations.} \label{fig:PSs}
\end{figure}

The absence of PDS\,70c in LB19+IB17+SB16 may perhaps reflect that the
signal detected in LB19 has been averaged with its absence in IB17. To
test this hypothesis we also imaged SB16+IB17, as shown in
Fig.\,\ref{fig:PDS70_concats_all}d, e and f. The noise level in
Fig.\,\ref{fig:PDS70_concats_all}e being 31.9\,$\mu$Jy\,beam$^{-1}$,
and since the data are aligned in position and flux, a 126\,$\mu$Jy
point source should have been picked up at 4$\sigma$. Instead, the
intensity at the location of PDS\,70c is $-$51\,$\mu$Jy\,beam$^{-1}$. In
order to test for synthesis imaging limitations due to the vicinity of
the strong and extended emission from the cavity wall, we injected 4
point sources in the residual visibility data, at the expected
location of PDS\,70c, and at symmetrical positions relative to the
disc axes. As summarised in Fig.\,\ref{fig:PSs}, the corresponding
restored image is consistent a 126\,$\mu$Jy point source at all 4
locations within 2\,$\sigma$.

%python summary_concats_IB17_withPS.py

The absence of PDS\,70c in IB17 suggests that it is a
variable source. \citet{Isella2019ApJ...879L..25I} reported the
detection of sub-mm continuum associated to PDS\,70c, using the same
dataset, but from their Fig.\,8 it seems to be offset northwards of
the H$\alpha$ peak. Also, their astrometry is tied to the central
continuum peak, which may be offset from the star (see below). In the
insets of Figs.\,\ref{fig:PDS70_concats_all}d and e there are indeed
filaments, or spurs, protruding away from the cavity edge more of less
at the position of the clump reported by
\citet{Isella2019ApJ...879L..25I}, but they do not match the position
of PDS\,70c given in \citet{Benisty2021ApJ...916L...2B}. In addition, these
spurs are roughly aligned with the elongated beam in IB17, and they
are absent in LB19, suggesting that they are imaging artefacts.

The non-detection of PDS\,70c in IB17
(Fig.\,\ref{fig:PDS70_concats_all}d) places a $3\sigma$ upper limit on
its flux density of 89\,$\mu$Jy. Since its flux in LB19 is
$(126\pm16.4)\,\mu$Jy\,beam$^{-1}$, if we assign the upper limit flux
density to PDS\,70c in IB17, then it brightened by $42\% \pm 13\% $ in
the interval of $\sim$1.75\,yr between IB17 and LB19. This might not
be so surprising since stellar, and probably also sub-stellar,
accretion is thought to be variable. Indeed,
\citet[][]{Szulagyi2020ApJ...902..126S} report a variability of the
H$\alpha$ flux from CPDs in the range 28\% to 58\%, over a time-scale
of $\sim 3$\,yr. Perhaps the signal stemming from PDS\,70c is due to
the free-free emission concomitant to the H$\alpha$ detection. If so,
its spectral index should be $\alpha \sim 0.7$
\citep[][]{WrightBarlow1975MNRAS.170...41W}, and should be more
conspicuous relative to the disc at lower frequencies.

Alternatively, the signal detected in LB19 might be due to dust being
heated up by episodic accretion, if it has time to heat-up in
1.75\,yr. Using the dust mass estimates from
\citet{Benisty2021ApJ...916L...2B} and
\citet{Portilla-Revelo2022A&A...658A..89P}, the Rosseland opacities
from \citet{Casassus2019MNRAS.486L..58C}, and for a flat surface
density out to 1\,au in radius, the vertical Rosseland optical depth
of this CPD would be in the range 5 to 5000, depending on maximum
grain size and temperature. The corresponding radiative diffusion
crossing time is at most 1\,month, so much shorter than the timescale
for the H$\alpha$ variability.

\subsubsection{The inner disc}

\begin{figure}
  \centering
  \includegraphics[width=\columnwidth,height=!]{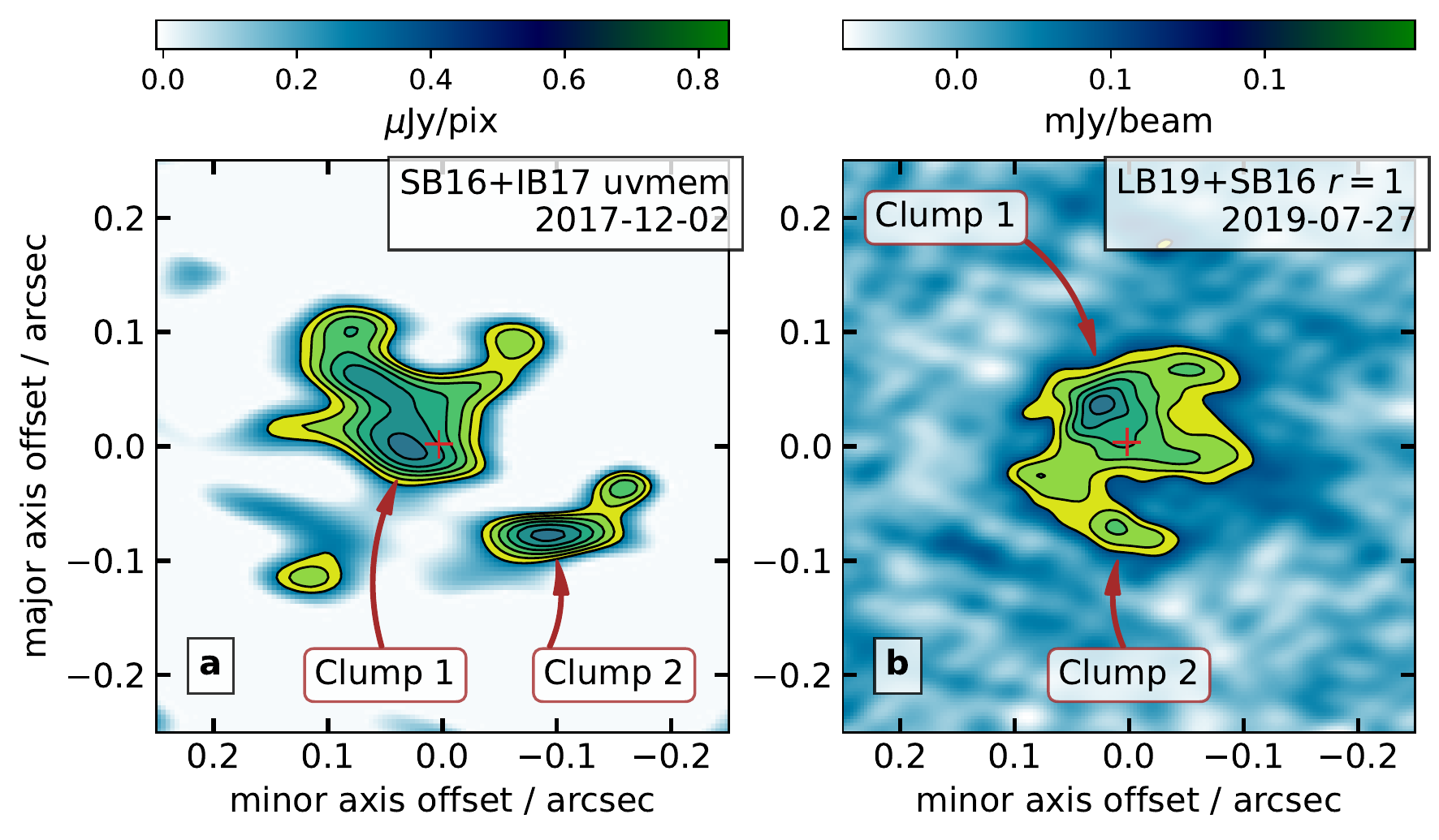}
  \caption{Face-on views of the inner disc in PDS\,70. Disc rotation
    is retrograde (clock-wise). The plus sign marks the origin and
    coincides with the ring center for each image. The $x$-axis and
    $y$-axis are aligned with the disc minor and major axis,
    respectively.  {\bf a)} Restored image for LB19+SB16 (from
    Fig.\,\ref{fig:PDS70_concats_all}b, with the same contours and noise
    levels). {\bf b)} {\sc uvmem} image from
    Fig.\,\ref{fig:PDS70_concats_all}f.} \label{fig:PDS70_cav}
\end{figure}

Another interesting feature of the available data in PDS\,70 is the
structure of the inner disc, which appears to be even more
statistically significant than the detection of PDS\,70c. The face-on
view in Fig.\,\ref{fig:PDS70_cav}b, obtained with the disc orientation
given above, shows extended substructure in the inner disc, with a
peak at $8\sigma$ (labeled `Clump\,1'in
Fig.\,\ref{fig:PDS70_cav}). Intriguingly, Clump\,1 is offset from the
ring center by $\sim0\farcs03$ to $0\farcs04$, which is comparable to
the natural-weights beam. At such small angular scales we expect
Keplerian rotation to result in detectable morphological changes in
the time-span of $\sim$1.75\,yr between IB17 and LB19. For a
$\sim$0.9\,$M_\odot$ star \citep[][]{Keppler2019A&A...625A.118K}, and
at a distance of 113.4\,pc, material at a radius of $0\farcs04$
(4.5\,au) would rotate by 62\,deg, or 95\,deg at $0\farcs03$
(3.4\,au). In Fig.\,\ref{fig:PDS70_cav}a we choose the SB16+IB17 {\sc
  uvmem} model image (from Fig.\ref{fig:PDS70_concats_all}f) for a
comparison with LB19, as it has a similar angular resolution as
SB16+LB19. Since the statistical properties of the model image are not
well constrained, the comparison is tentative. We find Clump\,1 offset
by $\sim$70\,deg in azimuth relative to SB16+LB19, as expected for
Keplerian rotation if the centre of mass coincides with the ring
center. The V-shaped structure with Clump\,1 at its vertex appears to
rotate as a rigid body, which might reflect a wave launched by
Clump\,1. However, in IB17 another peak, labeled Clump\,2 in
Fig.\,\ref{fig:PDS70_cav}, more compact and that reaches the same peak
intensity as Clump\,1, is found offset by $\sim0\farcs08$.  Keplerian
rotation on a circular orbit would be insufficient to bring Clump\,2
to its position in LB19; perhaps it is on a eccentric orbit and/or it
is rapidly infalling. If Clump\,2 is indeed the same structure in both
epochs, it is  fainter relative to Clump\,1 in LB19.

%~/common/ppdisks/PDS70/polarmaps/Mextract_profile_LB19SB16_offset.py
%~/common/ppdisks/PDS70/polarmaps/Mextract_profile_IB17SB16_modout_offset.py
%summary_cav_annot_offset.py

%summary_cav.py

%comparison between {\tt tclean} and
%{\tt uvmem} shown in Fig.\,\ref{fig:PDS70_briggs2} is the structure of
%the signal at the center of the disc. Both imaging strategies yield
%fairly similar structures, lending confidence in their reality.

The signal from the inner disc appears to be featureless in the images
from \citet[][their Figs.\,2 or 3 for Briggs r=1, or their Fig.\,6 for
Brigss $r=2$]{Benisty2021ApJ...916L...2B}. A possible explanation for
this difference could be that the structures picked-up here are part
of the residuals in \citet[][]{Benisty2021ApJ...916L...2B}. For
instance the bulk of the inner disc may have been fit by a
medium-scale Gaussian component in multi-scale {\tt tclean}, while the
details of the structure would have required deeper cleaning.  These
residuals would have been down-scaled by the JvM correction (see
Sec.\,\ref{sec:JvM}), and the inclusion of the extended multi-scale
{\tt tclean} component on this central emission may have resulted in a
seemingly featureless inner disc.

\section{Conclusions} \label{sec:conc}

In this article we have revisited the ALMA data on PDS\,70 using a
different imaging strategy than that in the original publications. We
used the synthesis imaging package {\sc uvmem}, which is an
alternative to {\tt tclean}. We have coupled {\sc uvmem} to the
standard self-calibration procedure implemented in CASA, within a
programming framework called {\sc OOselfcal}. We also proposed a
strategy for multi-epoch joint image restoration, which requires the
alignment of the visibility data. This alignment was performed in the
$uv$-plane and implemented in the {\sc VisAlign} package.

As a result of the application of {\sc VisAlign} and {\sc OOselfcal}
to the available data in PDS\,70, we have reached the following
conclusions:
\begin{enumerate}
\item The point source in PDS\,70c is detected, but only at the  8$\sigma$ level.
\item The same point source is absent in IB17, suggesting that it is variable by $42\% \pm 13\% $ on $1.75$\,yr timescales. 
\item The inner disc of PDS70 is structured and variable. Its peak is
  offset by $\sim 0\farcs04$ from the ring center. We tentatively
  observe  that the peak  appears to rotate according to Keplerian
  rotation relative to the ring center.
\end{enumerate}

We caution on the use of the JvM correction, which was applied in the
original publication of the long-baseline observations of PDS\,70. The
JvM correction exaggerates the peak-signal to noise ratio  of
the restored images by a factor of up to ten.

%we have summarised a procedure for image restoration,
%which is essentially standard, and thus does not involve novel
%definitions of the dirty beam as proposed in the recent literature
%(specifically in the so-called ``JvM correction''). A different dirty
%beam may exaggerate the PSNR 

\section*{Acknowledgements}
We thank the referee, Prof.\,St\'ephane Guilloteau, for a thorough
review and constructive comments that improved the manuscript, as well
as for fixing an error in the extraction of the power spectra in
Figs.\,\ref{fig:powerspectra} and \ref{fig:powerspectraIB17}. We also
thank Jonathan Williams and Sebasti\'an P\'erez for readings of the
manuscripts and useful comments.  We acknowledge the help of the {\sc
  dask} and {\sc astropy} communities for their technical support in
the development of {\sc VisAlign} and {\sc pyralysis}. S.C. and
M.C. acknowledge support from Agencia Nacional de Investigaci\'on y
Desarrollo de Chile (ANID) given by FONDECYT Regular grants 1211496,
ANID PFCHA/DOCTORADO BECAS CHILE/2018-72190574 and ANID project Data
Observatory Foundation DO210001.  This paper makes use of the
following ALMA data: ADS/JAO.ALMA\#{\tt 2015.1.00888.S,
  2017.A.00006.S, 2018.A.00030.S}. ALMA is a partnership of ESO
(representing its member states), NSF (USA) and NINS (Japan), together
with NRC (Canada), MOST and ASIAA (Taiwan), and KASI (Republic of
Korea), in cooperation with the Republic of Chile. The Joint ALMA
Observatory is operated by ESO, AUI/NRAO and NAOJ.

%%%%%%%%%%%%%%%%%%%%%%%%%%%%%%%%%%%%%%%%%%%%%%%%%%
\section*{Data Availability}

The reduced ALMA data presented in this article are available upon
reasonable request to the corresponding author.
The original or else non-standard software packages 
underlying the analysis are available at the following URLs: {\sc
  MPolarMaps}
\citep[\url{https://github.com/simoncasassus/MPolarMaps},][]{Casassus2021MNRAS.507.3789C}, {\sc uvmem}
\citep[\url{https://github.com/miguelcarcamov/gpuvmem},][]{Carcamo2018A&C....22...16C}, {\sc pyralysis}
(\url{https://gitlab.com/miguelcarcamov/pyralysis}), {\sc VisAlign}
(\url{https://github.com/simoncasassus/VisAlign}), {\sc OOselfcal}
(\url{https://github.com/miguelcarcamov/objectoriented_selfcal}).



\bibliographystyle{mnras}
%\bibliography{/home/simon/common/texinputs/merged.bib}
%\bibliography{merged.bib}
%\bibliography{/home/simon/common/ppdisks/PDS70/report_synthesis/merged.bib}

\input{pds70var.bbl}

%\begin{thebibliography}{99}
%\end{thebibliography}

%%%%%%%%%%%%%%%%%%%%%%%%%%%%%%%%%%%%%%%%%%%%%%%%%%

%%%%%%%%%%%%%%%%% APPENDICES %%%%%%%%%%%%%%%%%%%%%

\appendix

%%%%%%%%%%%%%%%%%%%%%%%%%%%%%%%%%%%%%%%%%%%%%%%%%%

\section{Image restoration} \label{sec:IS}

\subsection{Motivation for image restoration, and  for a  reconsideration of  the  JvM correction}

The interpretation of radio interferometer observations usually
involves a model of the sky image that accounts for the visibility
data \citep[except in the direct measurement of power spectra,
e.g. ][]{Pearson2003ApJ...591..556P}. These models are often
parametric \citep[e.g.][]{EHT2019ApJ...875L...1E}. Such parametric
models describe the source with few parameters, and if the model is an
adequate representation of the source, their posterior distributions
can be used to assess uncertainties. However, in general the model
image is non-parametric. The traditional synthesis imaging algorithm
{\sc clean} uses a collection of compact sources, represented by
spikes or circular Gaussians, and distributed by matching-pursuit
\citep[][]{Hogbom1974A&AS...15..417H,
  RauCornwell2011A&A...532A..71R}. A variety of Bayesian image
synthesis algorithms, or more generally regularized maximum likelihood
optimizations of the model image, have been proposed. Some use a
regular grid to represent the model image, which is then referred to
as the `deconvolved image', as in the {\sc uvmem} package
\citep[][]{Casassus2006, Carcamo2018A&C....22...16C}, or in 'Closure
Imaging' \citep[][]{Chael2018ApJ...857...23C,
  EHTimaging2019ApJ...875L...4E}, while others use the Voronoi
tessellation \citep[][]{Cabrera2008} or compressed sensing
\citep[e.g.][]{Wiaux2009MNRAS.395.1733W}.

However, the model images by themselves usually lack noise estimates.
This is because the inverse problem in image synthesis, i.e. the
production of a model image from sparse Fourier data, is intrinsically
ill-posed. After gridding the visibility data (see
Sec.\,\ref{sec:gridding} below), the number of free parameters reach
around a million to cover continuously the $uv$-plane, and exceeds the
number of independent data points.  Estimates of the uncertainties in
the model image are therefore challenging \citep[but
see][]{SuttonWandelt2006ApJS..162..401S,Casassus2015ApJ...811...92C}.

As an alternative, to convey the uncertainties in synthesis imaging it
is customary to ``restore'' the model images
\citep[e.g.][]{ThompsonMoranSwenson2017isra.book.....T,
  Briggs1999ASPC..180..127B} by convolution with an elliptical
Gaussian, which represents the interferometer angular resolution, and
the addition of the ``dirty map'' of the residuals, which is
essentially the inverse Fourier transform of the gridded visibility
residuals. These residuals encapsulate both the thermal noise and
synthesis-imaging artefacts. They may also include faint signal that
was missed in the model.

However, there are several different definitions of the dirty map,
that correspond to different gridding and data weighting
strategies. This degree of freedom has led to a reconsideration of the
dirty map units used in standard packages, in the so-called `JvM
correction' \citep[][]{JvM1995AJ....110.2037J,
  Czekala2021ApJS..257....2C}, resulting in noise values that differ
from the standard definition in image restoration. The dirty map of
the residuals is then scaled by the ratio of the dirty and clean
beams, thereby artificially down-scaling the noise level by up to one
order of magnitude. This appendix serves as a caution on the JvM
correction, which is increasingly applied in the recent literature
\citep[e.g.][ and the series of articles derived from the MAPS ALMA
Large Programme]{Benisty2021ApJ...916L...2B,
  Andrews2021ApJ...916...51A, Oberg2021ApJS..257....1O}.

\subsection{Definition of the restored image}

We assume that  a model image $I_m(\vec{x})$, defined onto a regular grid,  has  been  obtained  so that the associated model visibilities $\left\{ V_m(\vec{u}_j) \right\}_{j=1}^P$, 
\begin{equation}
  V_m(\vec{u}_j) = \int  I_m(\vec{x}) e^{2\pi i \vec{u}_j \cdot \vec{x}} dx dy, \label{eq:Vm}
\end{equation}
match the observed visibility data
$\left\{ V(\vec{u}_j) \right\}_{j=1}^P$ according to some criterion
(for instance a least-squares fit). To simplify notation we have
omitted from Eq.\,\ref{eq:Vm} the modulation by the primary beam and
by the Jacobian of the direction cosines (which are both close to
unity in the application to PDS\,70c). The residual visibilities are
$V_R(\vec{u}_j) = V(\vec{u}_j) - V_m(\vec{u}_j)$.
Given a model image  and a telescope angular resolution represented by an elliptical Gaussian $g_b(\vec{x})$, the restored image for a single pointing is
\begin{equation}
  I_R = I_m \ast g_b + R_D ,  \label{eq:IR}
\end{equation}
where $R_D(\vec{x})$ is the dirty map of the gridded residual
visibilities, $ \left\{ \tilde{V}_R(\vec{u}_k) \right\}_{k=1}^N$,
i.e. after averaging the sparsely sampled $uv$-data into a regular
grid with $N$ cells and weights $\{W_k\}_{k=1}^{N}$.  For an
extension of Eq.\,\ref{eq:IR} to mosaics see, for example,
\citet{CasassusROPH22021MNRAS.502..589C}.

\subsection{Gridding} \label{sec:gridding}

Since the visibility data correspond to the Fourier transform of the
sky image, a first approximation of the sky signal can be obtain by
averaging the visibility data into a regular grid and performing
the inverse Fourier transform. When this gridding procedure does not
extrapolate to the $uv$-cells devoid of data, which are assigned zero
weight, the result is called a dirty map.

There are several techniques for visibility gridding
\citep[e.g.][]{ThompsonMoranSwenson2017isra.book.....T,
  Briggs1999ASPC..180..127B}. Its simplest version is a
straight-forward application of counts in cells to grid the visibility
data $\{V_j\}_{j=1}^{P}$ in a regular grid in the $uv$-plane, with
$N=n\times m$ cells (for a model image defined on an $n\times m$
grid):
\begin{equation}
%\tilde{V}_k = \frac{\sum_{i=1}^{N_k} \omega_{{\rm nat},i} V_i}{\sum_{i=1}^{N_k}  \omega_{{\rm nat},i}}, \label{eq:gridV}
\tilde{V}_k = \frac{\sum_{i=1}^{N_k} w_i V_i}{\sum_{i=1}^{N_k}  w_i}, \label{eq:gridV}
\end{equation}
where $N_k$ is the number of visibilities that fall in cell $(u_k,v_k)$ and  $P = \sum_{k=1}^{N} N_k$.  The weights
$w_i$ are  propagated to the gridded visibilities, $\tilde{V}_k$, whose weights are then
\begin{equation}
  W_k = \sum_{i \in k} w_i. \label{eq:Wk}
\end{equation}

% or, more precisely, 
%\begin{align}
%    i =& (u_i/\Delta u) + n/2, \\
%    j =& (v_i/\Delta v) + m/2, \\
%    k =& n \times  i + j.
%\end{align}

There is a wide range of different choices for the visibility weights
$\{w_j\}_{j=1}^{P}$, with different units and resulting in different
intensity scales in the corresponding dirty maps (see
Sec.\,\ref{sec:dirtymap}). Natural weights
\citep[e.g.][]{ThompsonMoranSwenson2017isra.book.....T,
  Briggs1999ASPC..180..127B} assign the usual thermal noise weight to
each visibility datum:
\begin{equation}
    w_i = \omega_{{\rm nat},i} = \frac{1}{\sigma_i^2}. \label{eq:wadjust_nat}
\end{equation}
Other  weighting schemes are often used to control the effective interferometer resolution. In uniform weights \citep[e.g.][]{ThompsonMoranSwenson2017isra.book.....T,
  Briggs1999ASPC..180..127B} all cells are assigned the same weight, which is  achieved by redefining the weight for each datum:
\begin{equation}
    w_i = \frac{\omega_{{\rm nat},i}}{W_k}, \label{eq:wadjust_unif}
\end{equation}
where $W_k$ is the natural weight of the $k$-th cell where visibility $(u_i,v_i)$ falls. For uniform weights, Eq.\,\ref{eq:Wk} gives the weight of each $uv$-cell:
\begin{equation}
  W_k = \sum_{i \in k} w_i \equiv 1.
\end{equation}
In Briggs weighting \citep{briggs-thesis},
\begin{eqnarray}
    w_i &=& \frac{\omega_{{\rm nat},i}}{1+W_k f^2} \label{eq:wadjust_briggs}\\ 
    f^2 &=& \frac{(5 \cdot 10 ^{-r})^2}{\frac{\sum_k W_k ^ 2}{\sum_i \omega_{{\rm nat},i}}}, \nonumber
\end{eqnarray}
where $r$ is the Briggs robustness parameter ($r=2$ is almost
equivalent to natural weights, $r=-2$ is close to uniform weights).

The gridded weights $W_k$ derive from Eq.\,\ref{eq:Wk} using the
weights $w_i$, adjusted from the original weights
$\omega_{{\rm nat},i}$ according to Eqs.\,\ref{eq:wadjust_nat},
\ref{eq:wadjust_unif} or \ref{eq:wadjust_briggs}. Thus the units for
$W_k$ are different for each scheme.

\subsection{Dirty map, Dirty beam} \label{sec:dirtymap}

%
%regular grid with $N$ cells.

%Given the new proposals for image restoration in the recent literature, this Section gives details on the definitions used here. 

The dirty map might be thought of as a uniquely defined image, which would be the Fourier pair of the gridded visibilities. However, the degree of freedom in the choice of weighting scheme leads to different units for the dirty  map. We therefore adopt the following definition for the  dirty map:
\begin{equation}
    I_D(x,y) = \sum_k \alpha W_k \tilde{V}_k e^{-2 \pi i (ux+vy)} \Delta u \Delta v, \label{eq:ID0}
\end{equation}
where $\alpha$ is a constant that sets the units for $I_D$. In order
to determine $\alpha$ we consider a point source with flux density $F$ at
the phase center. We extract its flux from the gridded visibilities
 using a least-squares fit,
\begin{equation}
  \chi^2 = \sum_k (\tilde{V}_k  - \tilde{V}^m_k)^* (\tilde{V}_k  - \tilde{V}^m_k) W_k,
\end{equation}
where  $\tilde{V}^m_k = F$. The root of $\frac{\partial \chi^2 }{\partial F } = 0$ is 
\begin{equation}
  F = \frac{1}{2} \frac{ \sum_k (V_\circ + V_\circ^*) W_k}{\sum_k W_k}. \label{eq:flux0}
\end{equation}
Assuming that each visibility datum is independent, the error on $F$ is
\begin{equation}
  \sigma_F = \frac{1}{\sqrt{\sum_k W_k}}.
\end{equation}
On the other hand, the uncertainty on $I_D(\vec{x}=\vec{0})$ from Eq.\,\ref{eq:ID0} is
\begin{equation}
  \sigma_I = \alpha \Delta u \Delta v \sqrt{ \sum_k W_k^2 \sigma_k^2} =  \alpha \Delta u \Delta v \sqrt{\sum_k W_k}. 
\end{equation}

Next we assume that the $uv$-coverage is sufficient to use the dirty
map to approximate the sky image, smoothed by an elliptical Gaussian
that represents the image resolution. This requires that the
distribution of weights in the $uv$-plane is approximately Gaussian,
so that its Fourier pair is also an elliptical Gaussian in the image
plane, the so-called ``clean'' beam $g_b$ with
full-width-at-half-maximum major and minor axes ${\rm BMAJ}$ and
${\rm BMIN}$, and 
that subtends a solid angle
$\Omega_b  = \frac{\pi}{4\log(2)} {\rm BMAJ} \times {\rm BMIN}$. Given the degree of freedom in the weighting scheme, there
is no uniquely defined clean beam for a given visibility dataset, and
neither is there a unique angular resolution. In fact, even for the
same weighting scheme, different synthesis imaging packages do not
yield the same angular resolution.

%A common strategy to define the
%clean beam is to perform an elliptical Gaussian fit to the central
%peak of the dirty beam, and the resulting clean beam depends on
%implementation details.

If the units for $I_D$ are set to Jy$\,\Omega_b^{-1}$, then for a point source at the phase center, with flux density $F$, we should have
\begin{equation}
  \frac{\sigma_I}{ [ {\rm Jy}\,\Omega_b^{-1} ]  } = \sigma_F \label{eq:alpha}
\end{equation}
We can then solve for $\alpha$:
\begin{equation}
  \alpha = \frac{1}{\sum W_k \Delta u \Delta v}.
\end{equation}
The dirty map, in units of Jy$\,\Omega_b^{-1}$, is therefore:
\begin{equation}
  I_D(x,y) = \sum_k  \frac{W_k}{\sum_l W_l} \tilde{V}_k e^{2 \pi i (ux+vy)}, \label{eq:ID1}
\end{equation}
as is standard.  As an example, we consider the dirty map of a unit
spike at the phase center, which corresponds to the dirty-map
response, or point-spread function (PSF), and is usually called the
``dirty beam'':
\begin{equation}
  B(x,y) = \sum_k   \frac{W_k}{\sum_l W_l}   e^{2 \pi i (ux+vy)}.
\end{equation}
  
As another example, we consider the case of a perfect instrument with
uniform $uv$-coverage, that fills the $uv$-grid, with arbitrarily deep
data and constant weights $W_k$. In this case the dirty beam should be
a unit spike, and rather than an elliptical clean beam we would use
the solid angle subtended by a pixel in the sky image,
$\Omega_b = \Delta x \Delta y$, so that Eq.\,\ref{eq:ID1} yields the
same spike in units of Jy\,pix$^{-1}$.

% \footnote{here the parametric fit of a point source, but could also be a model  image}

We have checked that the synthesis imaging packages {\sc
  casa}\citep[][]{CASA2007ASPC..376..127M} and {\sc difmap}
\citep[][]{DIFMAP1997ASPC..125...77S} produce dirty maps with
consistent units. This can be confirmed by simulating visibility data
onto a unit spike in the absence of noise, and deriving the
corresponding dirty map (using task {\tt tclean} in {\sc casa} or {\tt
  mapplot} in {\sc difmap}), which yields peak intensities of
unity. In fact the dirty map of a unit spike is of course just the
dirty-map PSF and is one of the default products of tasks {\tt tclean}
and {\tt mapplot}.

\subsection{The JvM correction} \label{sec:JvM}

For consistency with the flux scale set by the visibility data
(Eq.\,\ref{eq:flux0}), the dirty map, as defined by
Eq.$\,\ref{eq:ID1}$, bears units of Jy\,$\Omega_b^{-1}$, where
$\Omega_b$ is the ``clean'' beam. By contrast, in the JvM correction
\citep[][]{JvM1995AJ....110.2037J, Czekala2021ApJS..257....2C}, a
`dirty beam' solid angle, here denoted by $\Omega_{\rm D}$, is introduced to
represent the solid angle subtended by the dirty map of a point
source, and to set the units of the dirty map in Eq.\,\ref{eq:ID1} to
Jy\,$\Omega_{\rm D}^{-1}$.  In this context, for image restoration the
residual dirty map must be scaled by the beam ratio
$\epsilon = $Jy\,$\Omega_b^{-1}$/Jy\,$\Omega_{\rm D}^{-1}$ before addition
with the convolved model image.

However the solid angle integral for $\Omega_{\rm D}$ will in
general not converge, as acknowledged by
\citet[][]{Czekala2021ApJS..257....2C}, and which can readily be
concluded from Fig.\,1 in \citep[][]{JvM1995AJ....110.2037J}. These
authors propose instead two different solutions to estimate
$\epsilon$.  In the original formulation of the JvM correction,
\citet[][]{JvM1995AJ....110.2037J} solve for $\epsilon$ by comparing
the total flux density extracted in a region between two versions of
the restored images. These two versions can be different depths of
cleaning, or simply the dirty and clean maps. However, the convolution
with dirty beam sidelobes will bias the flux density, in ways that
depend on the flux extraction aperture. In turn,
\citet[][]{Czekala2021ApJS..257....2C} adopt a different definition
for $\Omega_{\rm D}$ and $\epsilon$. They propose to perform an
azimuthal average of the dirty beam, and truncate the radial domain of
the dirty beam solid angle integral to the first null in the radial
direction.

In both versions of the JvM correction the values for $\epsilon$ range
from 1 to 1/10, and the extreme corresponds to natural weights.  But
both choices seem arbitrary, either in the definition of the flux
extraction aperture, or in the choice of domain for the solid
integral. Neither options are consistent with the requirement set by
Eq.\,\ref{eq:alpha}, since they both re-scale the units for the dirty
maps. The result is an artificial increase of sensitivity by up to a
factor of $1/\epsilon$.

The original motivation of the JvM correction may perhaps stem from
the normalisation of the dirty beam: since the dirty beam is the dirty
map response to a unit spike, it may be thought that its peak
intensity should correspond to the point source flux for the `correct
beam solid angle'. But this argument breaks down if the integrated
flux diverges.

% Don't change these lines
\bsp	% typesetting comment
\label{lastpage}
\end{document}